\documentclass[a4paper]{article}

\usepackage[utf8]{inputenc}
\usepackage{authblk}
\usepackage{amsmath}
\usepackage{amsthm}
\usepackage{bbold}
\usepackage{enumitem}
\usepackage[autosize]{dot2texi}
\usepackage{tikz}
\usetikzlibrary{shapes,arrows}
\usepackage{subcaption}
\usepackage{xcolor}
\usepackage[toc]{appendix}
\usepackage{hyperref}
\usepackage{graphicx}
\usepackage{floatpag}

\title{Decomposing the Jaccard Distance and the Jaccard Index in ABCDE}
\author[1]{Stephan van Staden}
\affil[1]{Google Switzerland GmbH}
\date{September 2024}

\newcommand{\Base}{\mathit{Base}}
\newcommand{\Exp}{\mathit{Exp}}
\newcommand{\Ideal}{\mathit{Ideal}}
\newcommand{\weight}{\mathit{weight}}
\newcommand{\JaccardDistance}{\mathit{JaccardDistance}}
\newcommand{\JaccardIndex}{\mathit{JaccardIndex}}
\newcommand{\SplitRate}{\mathit{SplitRate}}
\newcommand{\MergeRate}{\mathit{MergeRate}}
\newcommand{\Precision}{\mathit{Precision}}

\newcommand{\GoodMergeRate}{\mathit{GoodMergeRate}}

\newcommand{\DeltaPrecision}{\Delta\mathit{Precision}}
\newcommand{\DeltaRecall}{\Delta\mathit{Recall}}

\newcommand{\indicate}{\mathbb{1}}

\newcommand{\match}{\equiv}
\newcommand{\distinct}{\not\equiv}
\newcommand{\IQ}{\mathit{IQ}}

\newcommand{\AffectedItems}{\mathit{AffectedItems}}
\newcommand{\UnaffectedItems}{\mathit{UnaffectedItems}}

\newcommand{\SplitDistance}{\mathit{SplitDistance}}
\newcommand{\MergeDistance}{\mathit{MergeDistance}}
\newcommand{\GoodDistance}{\mathit{GoodDistance}}
\newcommand{\BadDistance}{\mathit{BadDistance}}
\newcommand{\GoodSplitDistance}{\mathit{GoodSplitDistance}}
\newcommand{\BadSplitDistance}{\mathit{BadSplitDistance}}
\newcommand{\GoodMergeDistance}{\mathit{GoodMergeDistance}}
\newcommand{\BadMergeDistance}{\mathit{BadMergeDistance}}
\newcommand{\GoodBadSplitMergeDistance}{\mathit{(Good|Bad)(Split|Merge)Distance}}
\newcommand{\GoodBadDistance}{\mathit{(Good|Bad)Distance}}
\newcommand{\GoodIndex}{\mathit{GoodIndex}}
\newcommand{\BadIndex}{\mathit{BadIndex}}
\newcommand{\AffectedJaccardIndex}{\mathit{AffectedJaccardIndex}}
\newcommand{\UnaffectedJaccardIndex}{\mathit{UnaffectedJaccardIndex}}
\newcommand{\AffectedGoodIndex}{\mathit{AffectedGoodIndex}}
\newcommand{\AffectedBadIndex}{\mathit{AffectedBadIndex}}

\begin{document}

\maketitle

\begin{abstract}
ABCDE is a sophisticated technique for evaluating differences between very large clusterings. Its main metric that characterizes the magnitude of the difference between two clusterings is the $\JaccardDistance$, which is a true distance metric in the space of all clusterings of a fixed set of (weighted) items. The $\JaccardIndex$ is the complementary metric that characterizes the similarity of two clusterings. Its relationship with the $\JaccardDistance$ is simple: $\JaccardDistance + \JaccardIndex = 1$.
This paper decomposes the $\JaccardDistance$ and the $\JaccardIndex$ further. In each case, the decomposition yields Impact and Quality metrics. The Impact metrics measure aspects of the magnitude of the clustering diff, while Quality metrics use human judgements to measure how much the clustering diff improves the quality of the clustering. The decompositions of this paper offer more and deeper insight into a clustering change. They also unlock new techniques for debugging and exploring the nature of the clustering diff. The new metrics are mathematically well-behaved and they are interrelated via simple equations. While the work can be seen as an alternative formal framework for ABCDE, we prefer to view it as complementary. It certainly offers a different perspective on the magnitude and the quality of a clustering change, and users can use whatever they want from each approach to gain more insight into a change.
\end{abstract}

{\bf Keywords:} Clustering evaluation, Clustering metrics, Clustering quality, Jaccard Distance, Jaccard Index, ABCDE

\section{Introduction}

ABCDE~\cite{vanstadengrubb2024abcde} is a sophisticated technique for evaluating differences in clusterings, where each clustering groups billions of items into clusters, and each item is associated with a positive $\weight$ which encodes its importance for the application at hand. Given a baseline clustering $\Base$ and an experiment clustering $\Exp$, ABCDE produces Impact metrics, which characterize the magnitude of the diff between $\Base$ and $\Exp$, and Quality metrics, which characterize the quality of the diff between $\Base$ and $\Exp$.

The original paper~\cite{vanstadengrubb2024abcde} that introduced ABCDE defined the $\JaccardDistance$ between $\Base$ and $\Exp$. The $\JaccardDistance$ is one of the main Impact metrics (the others are the $\SplitRate$ and the $\MergeRate$). In fact the $\JaccardDistance$ is a true distance metric for the space of all clusterings of a fixed set of weighted items, as was proved in~\cite{vanstaden2024pointwise}.

It is sometimes desirable to understand the $\JaccardDistance$ in more depth. For example, if $\Base$ is the clustering of yesterday's item data, and $\Exp$ is the clustering of today's item data, then the $\JaccardDistance$ between them will measure the magnitude of the cluster membership differences that resulted from updating the data. If this measurement indicates a large diff, then we might want to understand in more depth what happened. The technique for exploring the $\JaccardDistance$ mentioned in Section 4.1.1 of~\cite{vanstadengrubb2024abcde} can help, but it has limitations because it operates on the level of individual items. In particular, it produces a sample of individual items that are representative of the overall $\JaccardDistance$, but using them as exemplars for debugging the root causes of the diffs can still be hard, for example when such an item is in a large cluster, and hundreds of items were split off and merged in. The problem is that the item experienced a lot of change, and it is not clear exactly what to focus on.

It can be much more useful for debugging to have pairs of items that are representative of the $\JaccardDistance$. Each pair was either 1) in the same cluster in $\Base$ but in different clusters in $\Exp$, or 2) in different clusters in $\Base$ but in the same cluster in $\Exp$. Focusing on one pair at a time, and understanding why the items were split or merged, is a much easier task. The original formulation of ABCDE does sample pairs of items, but that is done to estimate $\DeltaPrecision$. Instead, we would like a way to sample pairs of items for the $\JaccardDistance$, and this paper shows how it can be done. The sample can also be explored interactively, along the lines of Section 4.1.1 of~\cite{vanstadengrubb2024abcde}, to understand the interactions (splits or merges) of pairs with particular attributes (provenances, types, etc.) and their approximate contribution to the overall $\JaccardDistance$.

Decomposing the $\JaccardDistance$ is a surprisingly deep topic. There are decompositions that characterize the clustering diff's Impact as well as its Quality. We can also decompose the $\JaccardIndex$. The decompositions of the $\JaccardDistance$ and the $\JaccardIndex$ can help to provide an unbiased estimate of $\DeltaPrecision$, but potentially with a larger confidence interval than with the sampling of~\cite{vanstadengrubb2024abcde}, which is tailored specifically for $\DeltaPrecision$ measurements.

The result can be viewed as an alternative formal framework for ABCDE. It is an attractive framework because there are many quantities that can be measured, and many of them are related via simple equations. In practical terms it means we can get more and deeper insight into clustering changes, and that the metrics are well-behaved and more interrelated.

Of course the approaches do not contradict each other, and we can use whatever we want from each of them to get more insight into a clustering change. For example, if high $\Precision$ is very important for the application at hand, then we would typically want to keep the primary focus on estimating $\DeltaPrecision$ with the smallest possible confidence interval, and the sampling and estimation described in Section~5 of~\cite{vanstadengrubb2024abcde} is best suited for that. But we can still debug the $\JaccardDistance$ using pairs or items, and compute the $\SplitDistance$ and $\MergeDistance$ metrics that are defined in this paper.

\section{High-level overview}

This paper contains definitions that decompose the $\JaccardDistance$ in several ways. The main equation that characterizes the diff's magnitude is:
$$\JaccardDistance = \SplitDistance + \MergeDistance$$
and the main equation that characterizes the diff's quality is:
$$\JaccardDistance = \GoodDistance + \BadDistance$$

Moreover, each of the four distances mentioned on the right-side of the two equations can be decomposed further:
\begin{align*}
\SplitDistance &= \GoodSplitDistance + \BadSplitDistance \\
\MergeDistance &= \GoodMergeDistance + \BadMergeDistance \\
\GoodDistance &= \GoodSplitDistance + \GoodMergeDistance \\
\BadDistance &= \BadSplitDistance + \BadMergeDistance
\end{align*}

We can compute exact values for the $\JaccardDistance$, the $\SplitDistance$, and the $\MergeDistance$.

We can obtain unbiased estimates with confidence intervals for $$\GoodBadSplitMergeDistance$$ -- a shorthand for $\GoodSplitDistance$, $\BadSplitDistance$, etc. -- as well as for $\GoodBadDistance$. This can be done directly from the sampled item pairs and their human judgements with standard statistics. 

Then we turn to the $\JaccardIndex$, which has a simple relationship with the $\JaccardDistance$:
$$\JaccardIndex + \JaccardDistance = 1$$

We decompose the $\JaccardIndex$ into two components to characterize its magnitude:
$$
\JaccardIndex = \AffectedJaccardIndex + \UnaffectedJaccardIndex
$$

The first term is for the items that are affected by the clustering change, in the sense that their $\Base$ and $\Exp$ clusters are not the same. We further decompose it to characterize the quality:
$$
\AffectedJaccardIndex = \AffectedGoodIndex + \AffectedBadIndex
$$

We can obtain unbiased estimates with confidence intervals for $\AffectedGoodIndex$ and $\AffectedBadIndex$.

The human judgements that were used to obtain the above estimates are sufficient to get an unbiased estimate with confidence interval for $\DeltaPrecision$. Everything fits together nicely, with the only caveat that the $\DeltaPrecision$ estimate is treated as a bonus -- it is not the primary objective and it might have a larger confidence interval compared to an estimate obtained with the sampling described in Section~5 of~\cite{vanstadengrubb2024abcde}.

\section{Preliminaries}

The terminology and notation of this paper is aligned with that of~\cite{vanstadengrubb2024abcde}:

\begin{itemize}
\item A \textit{clustering} partitions a set of \textit{items} into \textit{clusters}.
\item $\Base$ and $\Exp$ are clusterings that partition the same set of items $T$ into clusters. $T$ is also referred to as the \textit{population} of items.
\item $i$ and $j$ range over items. We use $I$ to denote an arbitrary set of items.
\item $\Base(i)$ denotes the cluster, i.e. the set of items, that are in the same cluster as $i$ in $\Base$. Similar notation is used for $\Exp$. It is always the case that $i \in \Base(i)$ and $i \in \Exp(i)$.
\item $i \match j$ means that items $i$ and $j$ are truly equivalent. An ideal clustering would put them together in the same cluster. True equivalence can be judged by humans. A clustering algorithm tries to approximate that. The fact that some approximations are better than others leads to the problem of evaluation, and hence techniques like ABCDE.
\item $i \distinct j$ means that items $i$ and $j$ are not truly equivalent. An ideal clustering would put them in separate clusters.
\item Each item $i$ is associated with a positive real weight, denoted $\weight(i)$. The weight of an item encodes its relative importance in the application at hand. The weights play a central role in the ABCDE metrics. They help to reflect that some clustering wins/losses are more important than others.
\item $\weight(I)$ denotes the weight of the set of items $I$. It is simply the sum of the weights of the members of $I$.
\item Most of the metrics discussed in this paper, including the $\JaccardDistance$ and the $\JaccardIndex$, are pointwise metrics~\cite{vanstaden2024pointwise} that are defined for individual items and arbitrary sets of items. The $\JaccardDistance$ of item $i$ is denoted by $\JaccardDistance(i)$. The $\JaccardDistance$ of a set of items $I$ is denoted by $\JaccardDistance(I)$. The overall $\JaccardDistance$ is $\JaccardDistance(T)$. A pointwise metric is lifted from the level of individual items to the level of sets of items by using expected values, i.e. weighted averages. So $\JaccardDistance(I)$ is the $\weight$-weighted average of $\JaccardDistance(i)$ over all $i \in I$.
\item $\indicate(e)$ is the indicator function for the Boolean expression $e$: it is defined to be 1 if $e$ is true and 0 otherwise.
\end{itemize}

\subsection{Estimating a weighted sum from a weighted sample}\label{EstimatingWeightedSum}

Many of the formal manipulations below use a sampling approach to estimate a metric that is expressed as a weighted sum. On a high level, the estimation procedure has three steps:
\begin{itemize}
\item Take a weighted sample of the population.
\item Compute the average (mean) over the sample.
\item Multiply the average by the total weight of the population to get an estimate of the metric.
\end{itemize}

The reasoning behind the procedure is as follows.  Suppose we want to estimate the sum:
$$\sum_{x} w_x f(x)$$
To do that we can take a weighted sample of elements $x_i$, which would let us then compute the mean:
$$\hat{f}(x) = \frac{\sum_{x_i} f(x_i)}{N}$$
To get an estimate of the metric we originally cared about, we can just see that:
$$E[\hat{f}(x)] = \frac{\sum_{x} w_x f(x)}{\sum_{x} w_x }$$
$$\sum_{x} w_x f(x) = E[\hat{f}(x)] \sum_{x} w_x$$
So the original sum can be estimated by the mean over a weighted sample, multiplied by the total weight of the whole population of elements that we were sampling over.

\section{Decomposing the $\JaccardDistance$}

\subsection{Decomposing the $\JaccardDistance$ of individual items}

The $\JaccardDistance$ of an item $i$ is defined as:
\begin{align}
\JaccardDistance(i) = \frac{\weight(\Base(i) \setminus \Exp(i)) + \weight(\Exp(i) \setminus \Base(i))}{\weight(\Base(i) \cup \Exp(i))}
\end{align}

(The numerator is equal to $\weight([\Base(i) \cup \Exp(i)] \setminus [\Base(i) \cap \Exp(i)])$.)

The two terms in the sum immediately suggest the definitions of the $\SplitDistance$ and the $\MergeDistance$:
\begin{align}
\SplitDistance(i) &=  \frac{\weight(\Base(i) \setminus \Exp(i))}{\weight(\Base(i) \cup \Exp(i))} \\
\MergeDistance(i) &= \frac{\weight(\Exp(i) \setminus \Base(i))}{\weight(\Base(i) \cup \Exp(i))}
\end{align}

These definitions trivially satisfy:
\begin{align}
\JaccardDistance(i) = \SplitDistance(i) + \MergeDistance(i)
\end{align}

We can define the $\GoodSplitDistance(i)$, the component of the split distance that is good (i.e. that is not truly equivalent to the vantage point item $i$), and the $\BadSplitDistance(i)$, and provide analogous definitions for merges:
\begin{align}
\GoodSplitDistance(i) &= \sum_{j \in \Base(i) \setminus \Exp(i)} \frac{\weight(j)}{\weight(\Base(i) \cup \Exp(i))} \indicate\left(i \distinct j\right) \\
\BadSplitDistance(i) &= \sum_{j \in \Base(i) \setminus \Exp(i)} \frac{\weight(j)}{\weight(\Base(i) \cup \Exp(i))} \indicate\left(i \match j\right) \\
\GoodMergeDistance(i) &= \sum_{j \in \Exp(i) \setminus \Base(i)} \frac{\weight(j)}{\weight(\Base(i) \cup \Exp(i))} \indicate\left(i \match j\right) \\
\BadMergeDistance(i) &= \sum_{j \in \Exp(i) \setminus \Base(i)} \frac{\weight(j)}{\weight(\Base(i) \cup \Exp(i))} \indicate\left(i \distinct j\right)
\end{align}

These definitions satisfy:
\begin{align}
\SplitDistance(i) &= \GoodSplitDistance(i) + \BadSplitDistance(i) \\
\MergeDistance(i) &= \GoodMergeDistance(i) + \BadMergeDistance(i)
\end{align}

We also define:
\begin{align}
\GoodDistance(i) &= \GoodSplitDistance(i) + \GoodMergeDistance(i) \\
\BadDistance(i) &= \BadSplitDistance(i) + \BadMergeDistance(i)
\end{align}

So we have:
\begin{align}
\JaccardDistance(i) = \GoodDistance(i) + \BadDistance(i)
\end{align}

\subsection{Decomposing the overall $\JaccardDistance$}

The metrics defined in the previous section are normal pointwise metrics~\cite{vanstaden2024pointwise}, which can be lifted to apply to arbitrary sets of items in the standard way: the metric for a set is simply the expected value of a member, i.e. the weighted average metric of the members of the set. This can be used to obtain metrics for individual clusters, for example, and also for the overall level (i.e. the entire clustering) by considering the set of all items $T$.

For example, the overall $\JaccardDistance$ between $\Base$ and $\Exp$ is simply the weighted average $\JaccardDistance$ of all items, i.e. the expected value of the $\JaccardDistance$ of an item:
\begin{align}
\JaccardDistance(T) = \sum_{i \in T} \frac{\weight(i)}{\weight(T)} \JaccardDistance(i)
\end{align}

The overall $\SplitDistance$ is the expected $\SplitDistance$ of an item, and similarly for $\MergeDistance$:
\begin{align}
\SplitDistance(T) &= \sum_{i \in T} \frac{\weight(i)}{\weight(T)} \SplitDistance(i) \\
\MergeDistance(T) &= \sum_{i \in T} \frac{\weight(i)}{\weight(T)} \MergeDistance(i)
\end{align}

We do the same for all the other metrics to end up with definitions that satisfy the following equations on any and every level of granularity, ranging from singleton sets to the set of all items $T$:
\begin{align}
\JaccardDistance &= \SplitDistance + \MergeDistance \\
\JaccardDistance &= \GoodDistance + \BadDistance \\
\SplitDistance &= \GoodSplitDistance + \BadSplitDistance \\
\MergeDistance &= \GoodMergeDistance + \BadMergeDistance \\
\GoodDistance &= \GoodSplitDistance + \GoodMergeDistance \\
\BadDistance &= \BadSplitDistance + \BadMergeDistance
\end{align}

\subsection{Calculation and estimation}

We can compute the $\JaccardDistance$, $\SplitDistance$ and $\MergeDistance$ exactly for any level of granularity we are interested in (in practice this means on the level of individual items, $\Base$ clusters, $\Exp$ clusters, and overall).

The quality assessment involves human judgements and estimation. We want unbiased estimates with confidence intervals for the overall
$$\GoodBadSplitMergeDistance(T)$$
We explain how that can be done by considering in detail how to estimate $\GoodSplitDistance(T)$.

Recall that:
\begin{align}
\GoodSplitDistance(T) &= \sum_{i \in T} \frac{\weight(i)}{\weight(T)} \GoodSplitDistance(i) \\
&= \sum_{i \in T} \frac{\weight(i)}{\weight(T)} \sum_{j \in \Base(i) \setminus \Exp(i)} \frac{\weight(j)}{\weight(\Base(i) \cup \Exp(i))} \indicate\left(i \distinct j\right) \\
&= \sum_{(i, j) \in \mathrm{AllSplits}} \frac{\weight(i) \cdot \weight(j)}{\weight(T) \cdot \weight(\Base(i) \cup \Exp(i))} \indicate\left(i \distinct j\right)
\end{align}
where $\mathrm{AllSplits} = \{(i, j) | i \in T \land j \in \Base(i) \setminus \Exp(i)\}$.

So we can sample item pairs from $\mathrm{AllSplits}$, where the weight of a pair $(i, j)$ is given by
\begin{align}\label{definition_of_w_ij}
w_{ij} = \frac{\weight(i) \cdot \weight(j)}{\weight(T) \cdot \weight(\Base(i) \cup \Exp(i))}
\end{align}
compute the average of $\indicate\left(i \distinct j\right)$ for the sample, and multiply that by $\SplitDistance(T)$ to get an unbiased estimate of $\GoodSplitDistance(T)$. To see why $\SplitDistance(T)$ is the right multiplier, we need to multiply by (see the explanation in Section~\ref{EstimatingWeightedSum}):
\begin{align}
& \sum_{(i, j) \in \mathrm{AllSplits}} \frac{\weight(i) \cdot \weight(j)}{\weight(T) \cdot \weight(\Base(i) \cup \Exp(i))} \nonumber \\
& = \sum_{i \in T} \frac{\weight(i)}{\weight(T)} \sum_{j \in \Base(i) \setminus \Exp(i)} \frac{\weight(j)}{\weight(\Base(i) \cup \Exp(i))} \\
& = \sum_{i \in T} \frac{\weight(i)}{\weight(T)} \frac{\weight(\Base(i) \setminus \Exp(i))}{\weight(\Base(i) \cup \Exp(i))} \\
& = \sum_{i \in T} \frac{\weight(i)}{\weight(T)} \SplitDistance(i) \\
& = \SplitDistance(T)
\end{align}

Similarly, we can estimate $\BadSplitDistance(T)$ by computing the average of $\indicate\left(i \match j\right)$ for the sample and multiplying it by $\SplitDistance(T)$, for the same reasons as above. Equivalently, we can use the equation $$\BadSplitDistance(T) = \SplitDistance(T) - \GoodSplitDistance(T)$$

The situation is analogous for merges: to estimate $\GoodMergeRate(T)$, we sample pairs of items $(i, j)$ from $\mathrm{AllMerges} = \{(i, j) | i \in T \land j \in \Exp(i) \setminus \Base(i)\}$ using weights $w_{ij}$ (yes, the same formula as for the splits!), compute the average of $\indicate\left(i \match j\right)$ for the sample, and multiply that by $\MergeDistance(T)$ to obtain an unbiased estimate of $\GoodMergeDistance(T)$. We can also estimate $\BadMergeDistance(T)$ by computing the average of $\indicate\left(i \distinct j\right)$ for the sample and multiplying it by $\MergeDistance(T)$.

\subsection{Summary of the estimation so far}

The set of all diff pairs of items is the union of the two disjoint sets, namely all splits and all merges:
\begin{align}
\mathrm{AllDiffs} &= \mathrm{AllSplits} \cup \mathrm{AllMerges} \\
\mathrm{AllSplits} &= \{(i, j) | i \in T \land j \in \Base(i) \setminus \Exp(i)\} \\
\mathrm{AllMerges} &= \{(i, j) | i \in T \land j \in \Exp(i) \setminus \Base(i)\}
\end{align}

For each diff pair $(i, j)$, compute the weight
$$w_{ij} = \frac{\weight(i) \cdot \weight(j)}{\weight(T) \cdot \weight(\Base(i) \cup \Exp(i))}$$
and sample diff pairs with replacement\footnote{See Appendix~A of~\cite{vanstadengrubb2024abcde} for information about sampling with replacement at scale.} according to their weight $w_{ij}$ to obtain a multiset $\mathrm{SampledPairs}$. After getting human judgements for these pairs, we can estimate the various overall quality metrics:
\begin{itemize}
\item  For $\GoodSplitDistance(T)$, we compute the average of
   $\indicate\left(i \distinct j\right)$ for all pairs in $\mathrm{SampledPairs} \cap \mathrm{AllSplits}$, and multiply the result by $\SplitDistance(T)$.
\item  For $\BadSplitDistance(T)$, we compute the average of
   $\indicate\left(i \match j\right)$ for all pairs in $\mathrm{SampledPairs} \cap \mathrm{AllSplits}$, and multiply the result by $\SplitDistance(T)$.
\item  For $\GoodMergeDistance(T)$, we compute the average of
   $\indicate\left(i \match j\right)$ for all pairs in $\mathrm{SampledPairs} \cap \mathrm{AllMerges}$, and multiply the result by $\MergeDistance(T)$.
\item  For $\BadMergeDistance(T)$, we compute the average of
   $\indicate\left(i \distinct j\right)$ for all pairs in $\mathrm{SampledPairs} \cap \mathrm{AllMerges}$, and multiply the result by $\MergeDistance(T)$.
\item  $\GoodDistance(T) = \GoodSplitDistance(T) + \GoodMergeDistance(T)$
\item  $\BadDistance(T) = \BadSplitDistance(T) + \BadMergeDistance(T)$
\end{itemize}

Note that we can easily compute a confidence interval for $\GoodDistance(T)$ because the sample for estimating $\GoodSplitDistance(T)$ is independent of the sample for estimating $\GoodMergeDistance(T)$ (the split pairs and the merge pairs are disjoint). Just use the fact that the standard error of a sum of independent random variables is the square root of the sum of the squares of the individual standard errors of the random variables. We can similarly compute a confidence interval for $\BadDistance(T)$.

\subsection{Sampling pairs of items for understanding the overall $\JaccardDistance$}

The goal is to sample a relatively small set of item pairs (say 5 million pairs) in a way we would if we wanted to estimate the overall $\JaccardDistance$. This small set is representative of the overall $\JaccardDistance$ and can be used to get a feel for the kinds of diffs that make up the $\JaccardDistance$. It can be explored interactively along the lines of Section~4.1.1 of~\cite{vanstadengrubb2024abcde}. For example, we would like to see how various slices of item pairs are affected by the clustering change, and drill deeper into slices we are curious about. And we can debug in detail a subsample of the diffs, or a subsample for a particular slice, to see exactly why and how the clustering diffs happened.

Recall that:
\begin{align}
& \JaccardDistance(T) \nonumber \\
& = \SplitDistance(T) + \MergeDistance(T) \\
& = \left(\sum_{(i, j) \in \mathrm{AllSplits}} w_{ij} \right) + \left(\sum_{(i, j) \in \mathrm{AllMerges}} w_{ij} \right) \\
& = \sum_{(i, j) \in \mathrm{AllDiffs}} w_{ij}
\end{align}

So we can obtain a sample in exactly the same way as for $\mathrm{SampledPairs}$ above -- in fact our sample could even be an extension to the one that is used there for human judgement. The draw count of a given pair divided by the overall draw count then indicates the relative importance of the pair. For example, if a given pair $(i, j)$ was drawn 5 times out of a total of $N$ draws, then for slicing and dicing purposes, $(i, j)$'s contribution to the overall $\JaccardDistance$ is $\frac{5}{N} \cdot \JaccardDistance(T)$. The sum of all contributions in the sample add up to the overall $\JaccardDistance$, and we can slice by applying a filter to the sampled pairs and summing up the contributions of the resulting pairs. We can also sum the contributions of the split pairs and the merge pairs separately to see the contributions of the slice to the overall $\SplitDistance$ and $\MergeDistance$ respectively.

\section{Decomposing the $\JaccardIndex$}

So far the discussion centered around the $\JaccardDistance$. We may well ask whether we can also estimate $\DeltaPrecision$ in this framework by sampling and judging additional pairs of items. $\Precision$, which measures the homogeneity of a cluster, also depends on the homogeneity of the part that is not split or merged. So the treatment in this section will go beyond decomposing the $\JaccardDistance$ -- it will also decompose the $\JaccardIndex$ -- and then a later section will get back to $\DeltaPrecision$.

\subsection{Decomposing the $\JaccardIndex$ of individual items}

The $\JaccardIndex$ from the vantage point of an item $i$ is defined as:
\begin{align}
\JaccardIndex(i) &= \frac{\weight(\Base(i) \cap \Exp(i))}{\weight(\Base(i) \cup \Exp(i))} \\
&= \sum_{j \in \Base(i) \cap \Exp(i)} \frac{\weight(j)}{\weight(\Base(i) \cup \Exp(i))}
\end{align}
It has a simple relationship with the $\JaccardDistance$:
\begin{align}
\JaccardIndex(i) + \JaccardDistance(i) = 1
\end{align}
We can decompose the $\JaccardIndex$ of an item $i$ as:
\begin{align}
\JaccardIndex(i) = \GoodIndex(i) + \BadIndex(i)
\end{align}

where

\begin{align}
\GoodIndex(i) &= \sum_{j \in \Base(i) \cap \Exp(i)} \frac{\weight(j)}{\weight(\Base(i) \cup \Exp(i))}  \indicate\left(i \match j\right) \\
\BadIndex(i) &= \sum_{j \in \Base(i) \cap \Exp(i)} \frac{\weight(j)}{\weight(\Base(i) \cup \Exp(i))}  \indicate\left(i \distinct j\right)
\end{align}

\subsection{Decomposing the overall $\JaccardIndex$}

The overall $\JaccardIndex$ is the expected $\JaccardIndex$ of an item:
\begin{align}
\JaccardIndex(T) &= \sum_{i \in T} \frac{\weight(i)}{\weight(T)} \JaccardIndex(i) \\
&= \sum_{i \in T} \sum_{j \in \Base(i) \cap \Exp(i)} \frac{\weight(i) \cdot \weight(j)}{\weight(T) \cdot \weight(\Base(i) \cup \Exp(i))} \\
&= \sum_{i \in T} \sum_{j \in \Base(i) \cap \Exp(i)} w_{ij}
\end{align}

(The last step holds because of equation~(\ref{definition_of_w_ij}) from before.)

To enable us to focus the human judgements only on the items that were affected by the clustering change, we partition the population of items $T$ into two disjoint sets\footnote{This was also done in Section~4.1.1 of~\cite{vanstadengrubb2024abcde}.}:

\begin{align}
\AffectedItems(T) &= \{i \in T | \Base(i) \neq \Exp(i)\} \\
\UnaffectedItems(T) &= \{i \in T | \Base(i) = \Exp(i)\}
\end{align}
and we define:
\begin{align}
\AffectedJaccardIndex(T) &= \sum_{i \in \AffectedItems(T)} \frac{\weight(i)}{\weight(T)} \JaccardIndex(i) \\
&= \sum_{i \in \AffectedItems(T)} \sum_{j \in \Base(i) \cap \Exp(i)} w_{ij} \\
\UnaffectedJaccardIndex(T) &= \sum_{i \in \UnaffectedItems(T)} \frac{\weight(i)}{\weight(T)} \JaccardIndex(i) \\
&= \sum_{i \in \UnaffectedItems(T)} \frac{\weight(i)}{\weight(T)} 1 \\
&= \frac{\weight(\UnaffectedItems(T))}{\weight(T)}
\end{align}
So the overall $\JaccardIndex$ has a simple decomposition:
\begin{align}
\JaccardIndex(T) = \AffectedJaccardIndex(T) + \UnaffectedJaccardIndex(T)
\end{align}

Please do not confuse $\AffectedJaccardIndex(T)$ with $\JaccardIndex(\AffectedItems(T))$. The latter is the expected $\JaccardIndex$ of an affected item, i.e. the weighted average $\JaccardIndex$ of the affected items. The two expressions are related by a simple equation that can be written in two ways:
\begin{align}
&\JaccardIndex(\AffectedItems(T)) = \frac{\weight(T)}{\weight(\AffectedItems(T))} \AffectedJaccardIndex(T) \\
&\AffectedJaccardIndex(T) = \frac{\weight(\AffectedItems(T))}{\weight(T)} \JaccardIndex(\AffectedItems(T))
\end{align}

We next define the overall $\AffectedGoodIndex$ as the expected $\GoodIndex$ of an affected item scaled by $$\frac{\weight(\AffectedItems(T))}{\weight(T)}$$ and similarly for the $\AffectedBadIndex$:

\begin{align}
& \AffectedGoodIndex(T) = \sum_{i \in \AffectedItems(T)} \frac{\weight(i)}{\weight(T)} \GoodIndex(i)  \\
&~~~~ = \sum_{i \in \AffectedItems(T)} \sum_{j \in \Base(i) \cap \Exp(i)} \frac{\weight(i) \cdot \weight(j)}{\weight(T) \cdot \weight(\Base(i) \cup \Exp(i))}  \indicate\left(i \match j\right) \\
&~~~~ = \sum_{i \in \AffectedItems(T)} \sum_{j \in \Base(i) \cap \Exp(i)} w_{ij}  \indicate\left(i \match j\right) \\
& \AffectedBadIndex(T) = \sum_{i \in \AffectedItems(T)} \sum_{j \in \Base(i) \cap \Exp(i)} w_{ij}  \indicate\left(i \distinct j\right)
\end{align}

So we can estimate $\AffectedGoodIndex(T)$ quite simply: sample pairs of items $(i, j)$, where $i \in \AffectedItems(T)$ and $j \in \Base(i) \cap \Exp(i)$, according to $w_{ij}$, compute the average of $\indicate\left(i \match j\right)$ for the sample, and multiply the result by $\AffectedJaccardIndex(T)$. For $\AffectedBadIndex(T)$, we multiply the sample's average of $\indicate\left(i \distinct j\right)$ with $\AffectedJaccardIndex(T)$, or equivalently, we can use $\AffectedBadIndex(T) = \AffectedJaccardIndex(T) - \AffectedGoodIndex(T)$.

For completeness, we state here the equations that hold on the overall level:
\begin{align}
&\JaccardIndex(T) + \JaccardDistance(T) = 1 \\
&\JaccardIndex(T) = \AffectedJaccardIndex(T) + \UnaffectedJaccardIndex(T) \\
&\AffectedJaccardIndex(T) = \AffectedGoodIndex(T) + \AffectedBadIndex(T)
\end{align}

\section{$\DeltaPrecision$ and its estimation}

\begin{figure*}[th!]
\centering
\begin{tikzpicture}[fill=gray]
\draw (-1.5,0) circle (2.5)
      (-1.5,2.5)  node [text=black,above] {$\Base(i)$}
      (1.5,0) circle (3)
      (1.5,3)  node [text=black,above] {$\Exp(i)$}
      (-0.4,-1.5) node [text=black] {$i$}
      (0,-2.75) circle (3.5)
      (0,-6.25) node [text=black,below] {$\Ideal(i)$}
      (-0.25,0) node [text=black] {\footnotesize $\GoodIndex$}
      (-0.25,1.2) node [text=black] {\footnotesize $\BadIndex$}
      (-2.7,0.6) node [text=black] {\footnotesize $\GoodSplitDistance$}
      (-2.18,-1.63) node [text=black] {\footnotesize $\BadSplitDistance$}
      (2.6,0.6) node [text=black] {\footnotesize $\BadMergeDistance$}
      (1.86,-1.6) node [text=black] {\footnotesize $\GoodMergeDistance$}
      ;
\end{tikzpicture}
\caption{The clustering quality situation from the perspective of item $i$. The item $i$ is always in the intersection of $\Base(i)$ and $\Exp(i)$ and $\Ideal(i)$, which is never empty. $\Ideal(i)$ is the set of all items that are truly equivalent to $i$. Each area inside the Venn diagram is labeled with its weight divided by $\weight(\Base(i) \cup \Exp(i))$. To save space we omit the suffix `$(i)$' from the labels. So, for example, the label $\GoodSplitDistance$ in the diagram stands for $\GoodSplitDistance(i)$.}\label{base_exp_ideal_cluster_diagram}
\end{figure*}
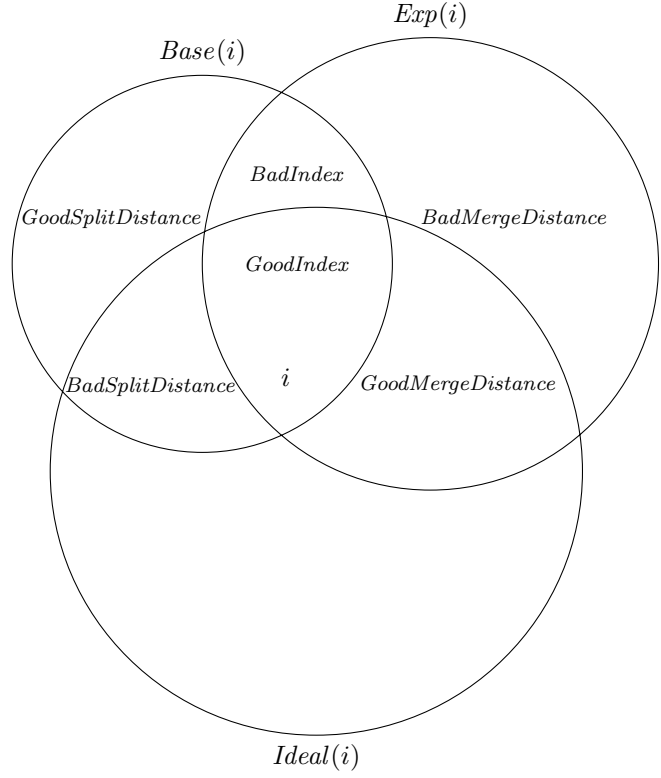

Figure~\ref{base_exp_ideal_cluster_diagram} shows the clustering quality situation from the perspective of an arbitrary item $i$.

From the diagram and the equations from before, it is easy to see that:
\begin{align}
\DeltaPrecision(i) =& \Precision_\Exp(i) - \Precision_\Base(i)  \\
=& \frac{\GoodIndex(i) + \GoodMergeDistance(i)}{\JaccardIndex(i) + \MergeDistance(i)} \nonumber \\
&-
\frac{\GoodIndex(i) + \BadSplitDistance(i)}{\JaccardIndex(i) + \SplitDistance(i)}
\end{align}

In this equation, notice that the terms align nicely vertically, i.e. $\GoodIndex$ is above $\JaccardIndex$, etc. Also, the denominators do not involve human judgements.

We can express the equation more simply by introducing abbreviations:
\begin{align}
m_i &= \JaccardIndex(i) + \MergeDistance(i) = \frac{\weight(\Exp(i))}{\weight(\Base(i) \cup \Exp(i))} \\
s_i &= \JaccardIndex(i) + \SplitDistance(i) = \frac{\weight(\Base(i))}{\weight(\Base(i) \cup \Exp(i))} \\
a_i &= \frac{s_i}{m_i s_i} = \frac{1}{m_i} = \frac{\weight(\Base(i) \cup \Exp(i))}{\weight(\Exp(i))} \\
b_i &= \frac{m_i}{m_i s_i} = \frac{1}{s_i} = \frac{\weight(\Base(i) \cup \Exp(i))}{\weight(\Base(i))}
\end{align}

Then the equation from before becomes:
\begin{align}
\DeltaPrecision(i) =&~ a_i \GoodIndex(i) + a_i \GoodMergeDistance(i) \nonumber \\
& - b_i \GoodIndex(i) - b_i \BadSplitDistance(i) \\
=&~ a_i \GoodMergeDistance(i) + (a_i - b_i) \GoodIndex(i) - b_i \BadSplitDistance(i) \\
=&
\left( a_i \sum_{j \in \Exp(i) \setminus \Base(i)} \frac{\weight(j)}{\weight(\Base(i) \cup \Exp(i))} \indicate\left(i \match j\right) \right) \nonumber \\
& +  
\left( (a_i - b_i) \sum_{j \in \Base(i) \cap \Exp(i)} \frac{\weight(j)}{\weight(\Base(i) \cup \Exp(i))}  \indicate\left(i \match j\right) \right) \nonumber \\
& +
\left( -b_i \sum_{j \in \Base(i) \setminus \Exp(i)} \frac{\weight(j)}{\weight(\Base(i) \cup \Exp(i))} \indicate\left(i \match j\right) \right)
\end{align}

We can express this as a standard weighted sum by introducing a label $l_{ij}$:
\begin{align}
\mathrm{if} \ j \in \Exp(i) \setminus \Base(i)&:\ \ l_{ij} = a_i  \\
\mathrm{if} \ j \in \Base(i) \setminus \Exp(i)&:\ \ l_{ij} = -b_i  \\
\mathrm{if} \ j \in \Base(i) \cap \Exp(i)&:\ \ l_{ij} = a_i - b_i
\end{align}

The $\DeltaPrecision$ from the vantage point of an item $i$ is then:
\begin{align}
\DeltaPrecision(i) = \sum_{j \in \Base(i) \cup \Exp(i)} \frac{\weight(j)}{\weight(\Base(i) \cup \Exp(i))} l_{ij} \indicate\left(i \match j\right)
\end{align}

Notice that, if $i \in \UnaffectedItems(T)$, then $\DeltaPrecision(i) = 0$, because if $\Base(i) = \Exp(i)$, then every $j \in \Base(i) \cup \Exp(i)$ will be in $\Base(i) \cap \Exp(i)$ and its label $l_{ij} = a_i - b_i$ will be zero.

Overall $\DeltaPrecision$ is then simply:
\begin{align}
& \DeltaPrecision(T) \nonumber \\
&= \sum_{i \in T} \frac{\weight(i)}{\weight(T)} \DeltaPrecision(i) \\
&= \left(\sum_{i \in \AffectedItems(T)} \frac{\weight(i)}{\weight(T)} \DeltaPrecision(i)\right) \nonumber \\
&~~~~ + \left(\sum_{i \in \UnaffectedItems(T)} \frac{\weight(i)}{\weight(T)} \DeltaPrecision(i)\right)\\
&= \sum_{i \in \AffectedItems(T)} \frac{\weight(i)}{\weight(T)} \DeltaPrecision(i) \\
&= \sum_{i \in \AffectedItems(T)} \sum_{j \in \Base(i) \cup \Exp(i)} \frac{\weight(i) \cdot \weight(j)}{\weight(T) \cdot \weight(\Base(i) \cup \Exp(i))} l_{ij} \indicate\left(i \match j\right) \\
&= \sum_{i \in \AffectedItems(T)} \sum_{j \in \Base(i) \cup \Exp(i)} w_{ij} l_{ij} \indicate\left(i \match j\right)
\end{align}

So we can estimate overall $\DeltaPrecision$ by sampling item pairs $(i, j)$, where $i \in \AffectedItems(T)$ and $j \in \Base(i) \cup \Exp(i)$, according to $w_{ij}$, computing the average of $l_{ij} \indicate\left(i \match j\right)$ for the sample, and multiplying that by
\begin{align}
& \sum_{i \in \AffectedItems(T)} \sum_{j \in \Base(i) \cup \Exp(i)} w_{ij} \nonumber \\
&= \left(\sum_{i \in \AffectedItems(T)} \sum_{j \in \Base(i) \setminus \Exp(i)} w_{ij}\right) + \left(\sum_{i \in \AffectedItems(T)} \sum_{j \in \Base(i) \cap \Exp(i)} w_{ij}\right) \nonumber \\
&~~~~ + \left(\sum_{i \in \AffectedItems(T)} \sum_{j \in \Exp(i) \setminus \Base(i)} w_{ij}\right) \\
&= \left(\sum_{i \in T} \sum_{j \in \Base(i) \setminus \Exp(i)} w_{ij}\right) + \left(\sum_{i \in \AffectedItems(T)} \sum_{j \in \Base(i) \cap \Exp(i)} w_{ij}\right) \nonumber \\
&~~~~ + \left(\sum_{i \in T} \sum_{j \in \Exp(i) \setminus \Base(i)} w_{ij}\right) \\
&= \left(\sum_{i \in T} \sum_{j \in \Base(i) \setminus \Exp(i)} w_{ij}\right) + \AffectedJaccardIndex(T) \nonumber \\
&~~~~ + \left(\sum_{i \in T} \sum_{j \in \Exp(i) \setminus \Base(i)} w_{ij}\right) \\
&= \left(\sum_{i \in T} \sum_{j \in \Base(i) \setminus \Exp(i)} \frac{\weight(i) \cdot \weight(j)}{\weight(T) \cdot \weight(\Base(i) \cup \Exp(i))}\right) + \AffectedJaccardIndex(T) \nonumber \\
&~~~~ + \left(\sum_{i \in T} \sum_{j \in \Exp(i) \setminus \Base(i)} \frac{\weight(i) \cdot \weight(j)}{\weight(T) \cdot \weight(\Base(i) \cup \Exp(i))}\right) \\
&= \left(\sum_{i \in T} \frac{\weight(i)}{\weight(T)} \sum_{j \in \Base(i) \setminus \Exp(i)} \frac{\weight(j)}{\weight(\Base(i) \cup \Exp(i))}\right) + \AffectedJaccardIndex(T) \nonumber \\
&~~~~ + \left(\sum_{i \in T} \frac{\weight(i)}{\weight(T)} \sum_{j \in \Exp(i) \setminus \Base(i)} \frac{\weight(j)}{\weight(\Base(i) \cup \Exp(i))}\right) \\
&= \left(\sum_{i \in T} \frac{\weight(i)}{\weight(T)} \SplitDistance(i)\right) + \AffectedJaccardIndex(T) \nonumber \\
&~~~~ + \left(\sum_{i \in T} \frac{\weight(i)}{\weight(T)} \MergeDistance(i)\right) \\
&= \SplitDistance(T) + \AffectedJaccardIndex(T) + \MergeDistance(T)
\end{align}

\subsection{Relationship with the estimation of $\DeltaPrecision$ in~\cite{vanstadengrubb2024abcde}}

The technique mentioned above for estimating $\DeltaPrecision$ can be viewed as an instance of importance sampling. While statistically sound, it can yield a less accurate/confident estimate of $\DeltaPrecision$ compared to the sampling and estimation technique of~\cite{vanstadengrubb2024abcde}, which is specifically tailored for tight estimates of $\DeltaPrecision$. In this paper the primary focus is not precision, but rather the $\JaccardDistance$ and the $\JaccardIndex$, and the estimate of $\DeltaPrecision$ should be seen as a bonus.

In the treatment of this paper, if $(i, j)$ is a merge pair, then
\begin{align}
w_{ij} &= \frac{\weight(i) \cdot \weight(j)}{\weight(T) \cdot \weight(\Base(i) \cup \Exp(i))} \\
l_{ij} &= \frac{\weight(\Base(i) \cup \Exp(i))}{\weight(\Exp(i))}
\end{align}
and in the treatment of~\cite{vanstadengrubb2024abcde}, the merge pair has:
\begin{align}
u_{ij} &= \frac{\weight(i)}{\weight(T)}\frac{\weight(j)}{\weight(\Exp(i))} \\
l_{ij} &= 1
\end{align}

The weight times the label is the same in both cases. The same is true for split pairs and intersection pairs. So we can view the absolute value of the label in this paper (which is equal to $u_{ij} / w_{ij}$) as the importance weight, i.e. a ``correction factor'' that compensates for the fact that the sampling distribution we used differs from the natural one for $\DeltaPrecision$. Importance weights make observations that were undersampled more important (and oversampled ones less important) in order to obtain a statistically sound estimate. Large importance weights, especially near-infinite ones, can lead to a large variance in the observations (because the effective sample size is small if it is dominated by a handful of very important observations) and hence to a large confidence interval of the estimate.

Generally speaking, one should use the natural sampling distribution of whatever estimate one wants to be the tightest, and then use the importance weighting technique to get other metrics that might be less important but still worth measuring. So if you consider $\DeltaPrecision$ to be primary, then you can apply the sampling technique of~\cite{vanstadengrubb2024abcde} and use importance sampling to obtain estimates for some metrics of this paper such as $\GoodSplitDistance$ etc.

There are limits to that approach, however. It is unfortunately not possible to obtain estimates for the quality decomposition of the $JaccardIndex$ in this way, because the sampling technique of~\cite{vanstadengrubb2024abcde} for $\DeltaPrecision$ will not sample any stable pairs of affected items for which the $\Base$ and the $\Exp$ clusters had exactly the same weight. Formally, if $(i, j) \in \mathrm{StablePairs}$, and $\weight(\Base(i)) = \weight(\Exp(i))$, then $u_{ij} = 0$, and hence the sampling technique of~\cite{vanstadengrubb2024abcde} will never include such pairs in the sample. Although they do not affect the value of $\DeltaPrecision$, they are needed to estimate $\AffectedGoodIndex(T)$ and $\AffectedBadIndex(T)$.

\section{Summary of the estimation}

We partition the set of all pairs of affected items
\begin{align*}
\mathrm{AllPairs} = \{(i, j) | i \in \AffectedItems(T) \land j \in \Base(i) \cup \Exp(i)\}
\end{align*}
of a clustering into 3 disjoint sets:
\begin{align*}
\mathrm{AllSplits} &= \{(i, j) | i \in \AffectedItems(T) \land j \in \Base(i) \setminus \Exp(i)\} \\
\mathrm{AllMerges} &= \{(i, j) | i \in \AffectedItems(T) \land j \in \Exp(i) \setminus \Base(i)\} \\
\mathrm{AllStablePairs} &= \{(i, j) | i \in \AffectedItems(T) \land j \in \Base(i) \cap \Exp(i)\}
\end{align*}
Each pair $(i, j) \in \mathrm{AllPairs}$ has a weight given by
$$w_{ij} = \frac{\weight(i) \cdot \weight(j)}{\weight(T) \cdot \weight(\Base(i) \cup \Exp(i))}$$
and a label $l_{ij}$ given by:
\begin{align}
\mathrm{if} \ j \in \mathrm{AllMerges}&:\ \ l_{ij} = a_i  \\
\mathrm{if} \ j \in \mathrm{AllSplits}&:\ \ l_{ij} = -b_i  \\
\mathrm{if} \ j \in \mathrm{AllStablePairs}&:\ \ l_{ij} = a_i - b_i \\
\end{align}
where
\begin{align*}
a_i &= \frac{\weight(\Base(i) \cup \Exp(i))}{\weight(\Exp(i))} \\
b_i &= \frac{\weight(\Base(i) \cup \Exp(i))}{\weight(\Base(i))}
\end{align*}

Sample pairs $(i, j)$ according to their weight $w_{ij}$ with replacement to obtain a multiset of sampled pairs $\mathrm{SampledPairs}$.

After obtaining human judgements for the sampled pairs, we can estimate the various overall quality metrics:
\begin{itemize}
\item
   For $\GoodSplitDistance(T)$, we compute the average of
   $\indicate\left(i \distinct j\right)$ for all pairs in $\mathrm{SampledPairs} \cap \mathrm{AllSplits}$, and multiply the result by $\SplitDistance(T)$.
\item
   For $\BadSplitDistance(T)$, we compute the average of
   $\indicate\left(i \match j\right)$ for all pairs in $\mathrm{SampledPairs} \cap \mathrm{AllSplits}$, and multiply the result by $\SplitDistance(T)$.
\item 
   For $\GoodMergeDistance(T)$, we compute the average of
   $\indicate\left(i \match j\right)$ for all pairs in $\mathrm{SampledPairs} \cap \mathrm{AllMerges}$, and multiply the result by $\MergeDistance(T)$.
\item 
   For $\BadMergeDistance(T)$, we compute the average of
   $\indicate\left(i \distinct j\right)$ for all pairs in $\mathrm{SampledPairs} \cap \mathrm{AllMerges}$, and multiply the result by $\MergeDistance(T)$.
\item
   $\GoodDistance(T) = \GoodSplitDistance(T) + \GoodMergeDistance(T)$
\item
   $\BadDistance(T) = \BadSplitDistance(T) + \BadMergeDistance(T)$
\item
   For $\AffectedGoodIndex(T)$, we compute the average of $\indicate\left(i \match j\right)$ for all pairs in $\mathrm{SampledPairs} \cap \mathrm{AllStablePairs}$, and multiply the result by $\AffectedJaccardIndex(T)$.
\item
   For $\AffectedBadIndex(T)$, we compute the average of $\indicate\left(i \distinct j\right)$ for all pairs in $\mathrm{SampledPairs} \cap \mathrm{AllStablePairs}$, and multiply the result by $\AffectedJaccardIndex(T)$.
\item
   For $\DeltaPrecision(T)$, we compute the average of
   $$l_{ij} \indicate\left(i \match j\right)$$
   over all $(i, j) \in \mathrm{SampledPairs}$, and multiply the result by $$\SplitDistance(T) + \AffectedJaccardIndex(T) + \MergeDistance(T)$$
\end{itemize}

We can easily compute a confidence interval for $\GoodDistance(T)$ as well as $\BadDistance(T)$, as mentioned before.

\subsection{Stratified sampling}

One concern with the sampling setup above is that it samples from a population that mixes the split/merge pairs and the stable pairs. So, for example, if we have a $\Base$ cluster with 1000 members and 1 item is split off into a cluster of its own, then from the vantage point of an item that remains (there are 999 of them), there is 1 split pair and 999 stable pairs (of which one is a self-pair), and from the vantage point of the item that is split off, there are 999 split pairs and 1 stable pair (which is a self-pair). So there are in total $999 + 999 = 1998$ split pairs and $999 \cdot 999 + 1 = 998002$ stable pairs (of which 1000 are self-pairs). If all the items have the same weight, then all the pairs will have equal weights, and the the population is dominated by stable pairs. Only $1998 / 1000000 = 0.1998\%$ are split pairs, and so the sampling would typically not get any or much information here for the $\mathit{(Good|Bad)SplitDistance}$.

To overcome this issue, one possibility is to use stratified sampling with two strata: the split/merge pairs, and the intersection pairs. By sampling more from the split/merge stratum compared to what we would sample without stratification, the $\GoodBadSplitMergeDistance$ can have tight estimates, while the $\mathit{Affected(Good|Bad)Index}$ quality estimates will be less tight with larger confidence intervals. Such a setup could be desirable in practice, but we leave the details for future work.

\subsection{More metrics}

The techniques of this paper provide a wealth of information about the clustering change. In particular, for the affected items, we can estimate:
\begin{itemize}
\item The expected good/bad split/merge distances.
\item The expected good/bad index.
\end{itemize}

These six quality metrics allow us to construct a summary diagram that shows the expected situation of an affected item. Together with basic information about the magnitude of the clustering change, such as the aggregate weights of the two sets of affected/unaffected items, we have everything we need to do back-of-the-envelope reasoning about the clustering change along the lines of~\cite{vanstaden2024clusteringqualitymetricsabcde}. Doing that can provide additional metrics, such as the overall $\DeltaRecall$ and the $\IQ$ of the change.

\section{Notes for implementers}

\begin{itemize}
\item The set $\mathrm{AllStablePairs}$ includes the set $$\mathrm{AllSelfPairs} = \{(i, i) | i \in \AffectedItems(T)\}$$ in which each affected item is paired with itself. So in the definition of $\mathrm{AllStablePairs}$ above, \textit{please do not naively assume} $i \neq j$.
\item The $\mathrm{SampledPairs}$ typically include many pairs from $\mathrm{AllSelfPairs}$ in practice. While we do not need to get these pairs judged by humans because an equivalence relation is always reflexive, it is very important to keep these samples around and to treat them all as if they got the human judgement $i \match j$.
\item The formula for $w_{ij}$ divides by $\weight(T)$ which is the same for every pair. We can omit that when sampling, since scaling every weight by a constant factor will not affect the sample, but we still have to include it in the multipliers for the metrics.
\item Sometimes humans are uncertain and cannot decide whether $i \match j$ or $i \distinct j$. Sometimes it is even impossible to ask the judgement question, for example when the data of $i$ and/or $j$ is not available anymore. In such cases it makes sense to exclude these sampled pairs from the metrics, which can be done by excluding them from $\mathrm{SampledPairs}$ in the formulas above. \\
\textbf{Caveat}: For $\DeltaPrecision(T)$, $\AffectedGoodIndex(T)$ and $\AffectedBadIndex(T)$, the remaining sampled pairs can be biased, because the sampled pairs that are also in $\mathrm{AllSelfPairs}$ will always remain (in practice they can easily comprise 30\% of all the sampled pairs and they always get the ``judgement" $i \match j$). This can be remedied by classifying the sampled pairs into classes, e.g. $\mathrm{SelfPairs}$, $\mathrm{SplitPairs}$, $\mathrm{MergePairs}$, $\mathrm{IntersectionPairs}$ (for $\mathrm{StablePairs} \setminus \mathrm{SelfPairs}$), and introducing weights for the remaining sampled pairs such that the total weight of the remaining pairs in each class is equal to the total weight of the originally sampled pairs in each class. For example, if we sampled 1000 split pairs, but only 800 have judgements, then each of the remaining split pairs will get a weight of 1000/800 = 1.25.
\item The technique of weighting the sampled pairs mentioned in the previous point can be applied to arbitrary classes/slices of sampled pairs: if it is hard to answer the judgement questions of some slices, then the weights can ensure that these slices do not get underrepresented in the metrics.
\item It is possible to report confidence intervals by computing the standard errors of the metrics. One very useful result is that $\mathrm{StdErr}(c \cdot X) = c \cdot \mathrm{StdErr}(X)$, which holds because $\mathrm{StdDev}(c \cdot X) = c \cdot \mathrm{StdDev}(X)$. Another useful result is that, if $X$ and $Y$ are derived from independent samples, then $\mathrm{StdErr}(X + Y) = \sqrt{[\mathrm{StdErr}(X)]^2 + [\mathrm{StdErr}(Y)]^2}$. Here are the details for the various metrics:
\begin{itemize}
  \item The standard error of $\GoodSplitDistance(T)$ is $\SplitDistance(T)$ times the standard error of $\indicate\left(i \distinct j\right)$ for all pairs in $\mathrm{SampledPairs} \cap \mathrm{AllSplits}$.
  \item  $\BadSplitDistance(T)$ has the same standard error as that of \\ $\GoodSplitDistance(T)$.
  \item The standard error of $\GoodMergeDistance(T)$ is $\MergeDistance(T)$ times the standard error of $\indicate\left(i \match j\right)$ for all pairs in $\mathrm{SampledPairs} \cap \mathrm{AllMerges}$.
  \item $\BadMergeDistance(T)$ has the same standard error as that of \\ $\GoodMergeDistance(T)$.
  \item The standard error of $\GoodDistance(T)$ is the square root of the sum of (the squared standard error of $\GoodSplitDistance(T)$) and (the squared standard error of $\GoodMergeDistance(T)$).
  \item $\BadDistance(T)$ has the same standard error as that of $\GoodDistance(T)$.
  \item The standard error of $\AffectedGoodIndex(T)$ is $\AffectedJaccardIndex(T)$ times the standard error of $\indicate\left(i \match j\right)$ for all pairs in $\mathrm{SampledPairs} \cap \mathrm{AllStablePairs}$. \\
  The removal of the sampled pairs without judgements and the subsequent weighting of the remaining pairs means that we have a weighted sample, whose standard error is discussed in Section~5.8 of~\cite{vanstadengrubb2024abcde}.
  \item $\AffectedBadIndex(T)$ has the same standard error as that of \\ $\AffectedGoodIndex(T)$.
  \item The standard error of $\DeltaPrecision(T)$ is $$\SplitDistance(T) + \AffectedJaccardIndex(T) + \MergeDistance(T)$$ times the standard error of  $$l_{ij} \indicate\left(i \match j\right)$$
   for all pairs in $\mathrm{SampledPairs}$. \\
   If there are item pairs without clear judgement verdicts, we can remove them and perform weighting of the remaining pairs as described before. If we do that, then we should use the formula for the standard error of a weighted sample, just as we did for $\AffectedGoodIndex(T)$.
\end{itemize}
Note that these confidence intervals quantify only the uncertainty inherent in the sampling. They do not quantify the uncertainty in the human judgements. It is possible to quantify that by replicating questions (i.e. asking multiple humans the same question) and using bootstrapping techniques, but that is impractical unless the budget for human judgements is large.
\item We recommend sampling with replacement. In practical applications the pairs can have a broad range of weights, and it is common to see pairs with a draw count greater than one. Sampling with replacement at scale is discussed at length in Appendix~A of~\cite{vanstadengrubb2024abcde}.
\end{itemize}

\section{Conclusion}
This paper decomposes the $\JaccardDistance$ and the $\JaccardIndex$ into Impact and Quality metrics. The goal is to obtain more and deeper insight into a clustering change. The metrics themselves are mathematically well-behaved and they are interrelated via simple equations. They also unlock new techniques for debugging and exploring the nature of the clustering diff. While the work can be seen as an alternative formal framework for ABCDE, we prefer to view it as complementary. It certainly offers a different perspective on the magnitude and the quality of a clustering change, and users are free to use whatever they want from each approach to get more insight into the change at hand.

\subsubsection*{Acknowledgements}
Many thanks to Alexander Grubb for extensive comments on earlier versions of this work.

\bibliographystyle{plain}
\bibliography{main}

\begin{thebibliography}{1}

\bibitem{vanstaden2024clusteringqualitymetricsabcde}
Stephan van Staden.
\newblock More clustering quality metrics for {ABCDE}, 2024.
\newblock \url{https://arxiv.org/abs/2409.13376}.

\bibitem{vanstaden2024pointwise}
Stephan van Staden.
\newblock Pointwise metrics for clustering evaluation, 2024.
\newblock \url{https://arxiv.org/abs/2405.10421}.

\bibitem{vanstadengrubb2024abcde}
Stephan van Staden and Alexander Grubb.
\newblock {ABCDE}: Application-based cluster diff evals, 2024.
\newblock \url{https://arxiv.org/abs/2407.21430}.

\end{thebibliography}

\end{document}